\documentclass[12pt]{article}

\usepackage[top=1in, bottom=1in, left=1in, right=1in]{geometry}
\usepackage{amsmath, amsfonts}
\usepackage{setspace}
\usepackage{mdframed}
\usepackage{graphicx}
\usepackage{subcaption}
\usepackage{natbib}   

\title{Fast permutation tests and related methods, for association between rare variants and binary outcomes}
\author{
        Arjun Sondhi  \\
        Department of Biostatistics\\
        University of Washington\\
        Seattle, WA, USA\\
            \and
        Kenneth M. Rice, PhD\\
        Department of Biostatistics\\
        University of Washington\\
        Seattle, WA, USA
}
\date{March 28, 2017}

\newcommand{\bm}{\mbox{\boldmath $m$}}
\newcommand{\br}{\mbox{\boldmath $r$}}

\begin{document}
\maketitle

\let\thefootnote\relax\footnotetext{This is the pre-peer reviewed version of the following article: Sondhi A, Rice KM. Fast permutation tests and related methods, for association between rare variants and binary outcomes. Ann Hum Genet. 2017;00:1--9, which has been published in final form at https://doi.org/10.1111/ahg.12229. This article may be used for non-commercial purposes in accordance with Wiley Terms and Conditions for Self-Archiving.}

\clearpage

\begin{abstract}
\noindent In large scale genetic association studies, a primary aim is to test for association between genetic variants and a  disease outcome. The variants of interest are often rare, and appear with low frequency among subjects. In this situation, statistical tests based on standard asymptotic results do not adequately control the Type I error rate, especially if the case:control ratio is unbalanced. In this paper, we propose the use of permutation and approximate unconditional tests for testing association with rare variants. We use novel analytical calculations to efficiently approximate the true Type I error rate under common study designs, and in numerical studies show that the proposed classes of tests significantly improve upon standard testing methods. We also illustrate our methods in data from a recent case-control study, for genetic causes of a severe side-effect of a common drug treatment.
\end{abstract}


\clearpage

\section{Introduction}

Association studies are often performed for binary traits, providing new knowledge of the genetic causes of human diseases, using data from case-control and cohort studies~\citep{GWAS1,GWAS2,GWAS3,GWAS4}. Recent advances in sequencing technology have made it practical to type essentially every variant on the genome; to avoid spurious findings, very low Type I error rates must therefore be maintained~\citep{GWASsig}. However, for the rare variants now being studied, standard analytic approaches do not reliably achieve their nominal rates,~\citep{standard2,standard1,Ma} and may permit too many Type I errors. The problem can be particularly severe when the ratio of cases to controls is extreme. Adjustments that maintain control at the nominal rate can be conservative, leading to loss of power relative to methods that control Type I errors more accurately.

In this paper, motivated by work in a case-control study of rhabdomyolosis, we develop methods with improved control of the Type I error rate, when testing single rare variants for association with binary traits. In Section~\ref{sec2}, we explain a novel numerical method that approximates the actual Type I error rate of a test statistic given sample size, significance level, and a variant's expected frequency; we also show how the same basic ideas can be used in permutation and approximate unconditional tests, and how the ideas can be used when adjusting for covariates. In Section~\ref{sec3}, we give the results of the numerical studies performed, demonstrating improvements over standard asymptotic tests. Section~\ref{sec4} applies these methods to data from a case-control study of statin-related rhabdomyolysis, and we conclude with a short discussion, including details of an R package that implements our methods. 

\section{Methods}
\label{sec2}

%\itshape
%\begin{enumerate}
%\item Given $P(G=1) = \frac{MAC}{N}$, obtain the range of plausible total minor allele carriers in a population of size $m_0 + m_1$. We defined plausible as having probability $ > 10^{-12} $ of being observed.
%\item Enumerate all the possible combinations of case and control carriers (denoted $r_1$ and $r_0$ respectively) that sum to each plausible total count.
%\item Calculate the probability of observing each dataset $(m_0, m_1, r_0, r_1)$ under the null hypothesis of $Y$-$G$ independence.
%\item Calculate a test statistic and associated p-value for each dataset.
%\item The average Type I error rate is the sum of the probabilities of datasets that result in p-values $< \alpha$.
%\end{enumerate} 
%\upshape

With rare variants, homozygotes are so rare as to be negligible for analysis, and it suffices to consider whether subjects have any copies of the variant present ($G=1$) or not ($G=0$). This step also means it is simple to enumerate all possible datasets; given fixed numbers of cases and controls ($m_1$ and $m_0$ respectively), we need only consider the number of cases ($r_1$) and controls ($r_0$) with the variant (adjustment for covariates is considered in Section 2.3.).

For a variant with a given minor allele frequency ($MAF$), following \citet{Ma}, we define expected number of minor allele carriers as $EMAC  = (m_0 + m_1) \times (1 - (1-MAF)^2)$. Under the null hypothesis of no association, it follows that
\begin{eqnarray*}
r_1 & \sim & Binom\bigg(m_1, \frac{EMAC}{m_0+m_1}\bigg) \\
r_0 & \sim & Binom\bigg(m_0, \frac{EMAC}{m_0+m_1}\bigg), 
\end{eqnarray*}
independently. Therefore, given a value of EMAC --- or equivalently MAF --- and the fixed numbers of cases and controls in the study at hand, we can simply write down the probability of seeing all possible datasets under the null hypothesis.

In theory, this direct enumeration allows exact calculation of the Type I error rate for any test: the Type I error rate is the sum of the probabilities of the datasets for which a significant test result is returned. Formally, a dataset $(m_0,m_1,r_0,r_1)$ returns a significant test result when its associated p-value $p(r_0, r_1; m_0, m_1) \le \alpha$, where $p(r_0, r_1; m_0, m_1) = \mathbb{P}[\,|T| \ge T_{obs} ; H_0 \,]$ is the probability of test statistic $T$ equaling or exceeding the observed value $T_{obs}$ when the null hypothesis holds. The Type I error rate of the test at nominal level $\alpha$ is then defined as
\begin{equation}
T1ER(\alpha) = \sum_{0\leq r_0\leq m_0, 0\leq r_1\leq m_1} f(r_0, r_1; m_0, m_1, EMAC)1_{p(r_0, r_1; m_0, m_1) < \alpha},
\label{eqn1}
\end{equation}
where $f(r_0, r_1; m_0, m_1, EMAC)$ denotes the probability of observing data $r_0, r_1$ under the null hypothesis, given $m_0$, $m_1$, and a specific EMAC. Although not discussed further, the approach is easily adapted to $p$-values that use lower tail areas -- below $T_{obs}$ instead of above, or two-sided tests that examine tail areas beyond $\pm T_{obs}$.

In practice, the sum of $(m_0+1)\times(m_1+1)$ terms in~(\ref{eqn1}) may be large, making computation too slow for some purposes. However, for work on rare variants, almost all of the summands contribute negligibly to the overall Type I error rate. A practical solution is therefore to truncate the summation in~(\ref{eqn1}) by zeroing-out terms that, in total, represent no more than a small fraction of the Type I error rate. 

\begin{figure}[h]
\centering
a) \raisebox{-3.2in}{\includegraphics[width=0.46\textwidth]{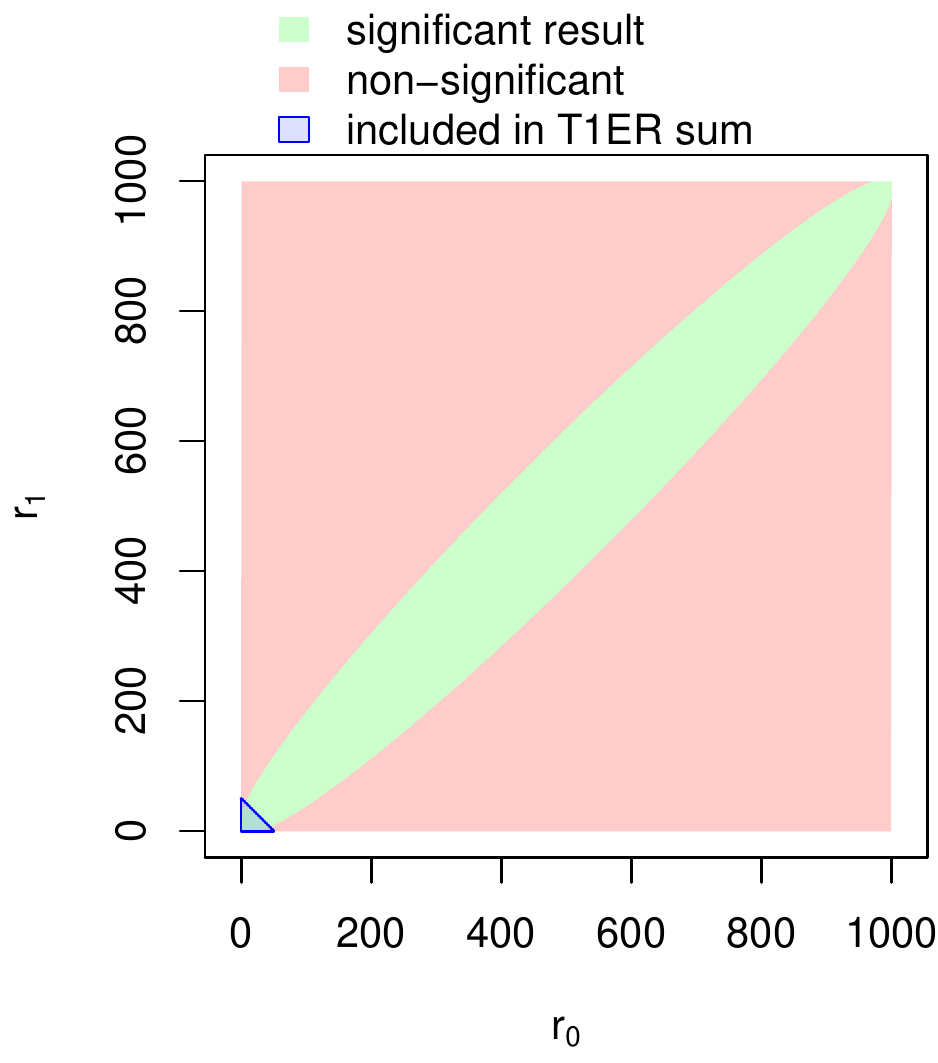}} b) \raisebox{-3.2in}{\includegraphics[width=0.46\textwidth]{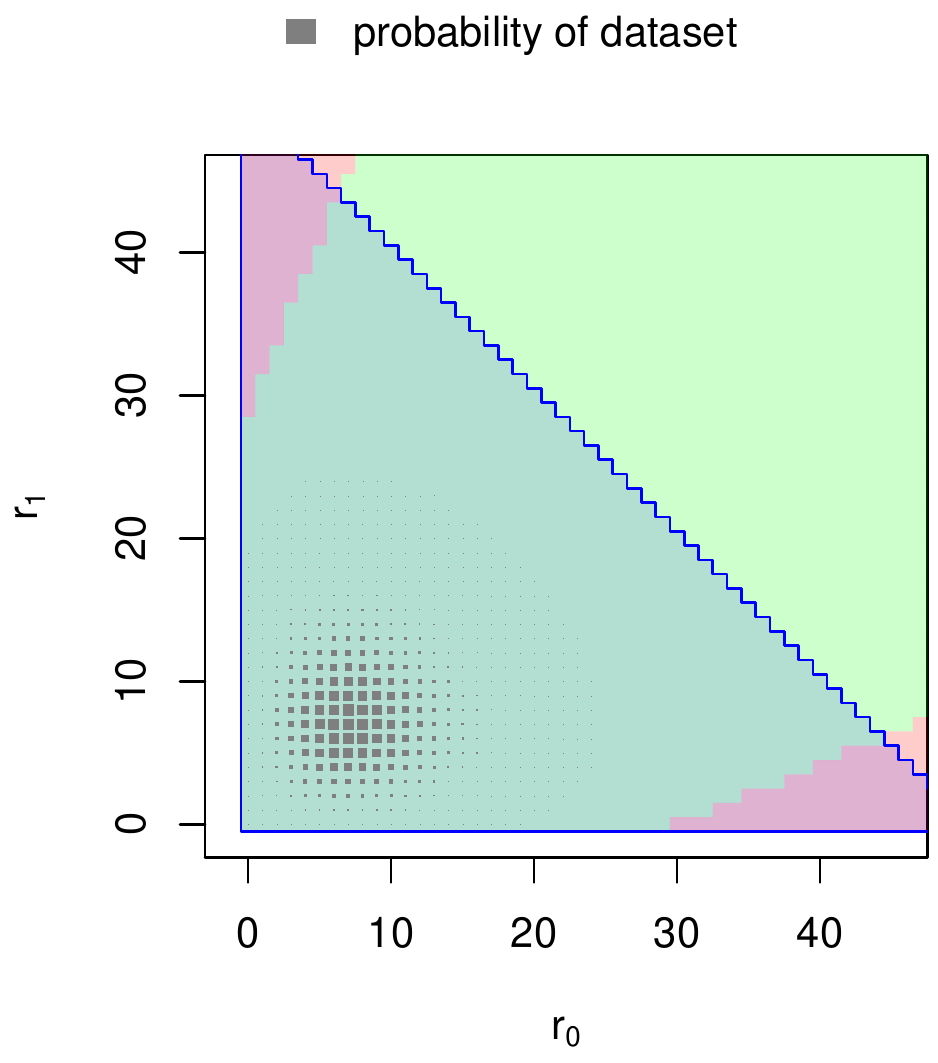}}
\caption{Possible datasets and their contribution to $T1ER$ for the standard Score test, for $m_0=m_1=1000$ and $EMAC=15$. In a), the red/green zones indicate datasets where the standard score test is significant/not-significant at nominal $\alpha=5\times 10^{-8}$; the blue zone shows terms that are not zeroed-out using the truncation described in Section~\ref{sec2}. In b) the same situation is shown, zoomed in and with box size proportional to the probability of each dataset; the actual $T1ER(\alpha)$ is given by the sum of the box areas in the two red zones. Zeroing-out contributions beyond the blue region, where $r_0+r_1> 50$, gives an approximation error in the $p$-value of no more than $10^{-12}$.}
\label{fig1}
\end{figure}

Taking this approach, in our work we will zero-out terms in~(\ref{eqn1}) representing datasets for which $r_0+r_1$ exceeds the upper $10^{-12}$ quantile of the distribution of $r_1 + r_0$, i.e. of $Binom(m_0+m_1, \frac{EMAC}{m_0+m_1})$ (Figure~\ref{fig1} gives a graphical description of this process). By setting these terms to zero, in this example, we therefore understate the Type I error rate by no more than $10^{-12}$, which is acceptable given our focus on Type I error rates near $\alpha=5\times 10^{-8}$, and maintain practical computation times, even for large studies. For example, performing this calculation for $(m_0, m_1, EMAC) = (500, 500, 15)$ at $\alpha = 5 \times 10^{-8}$ using the standard Score test takes 0.05 seconds on a standard laptop. Without the zeroing-out method, the calculation takes 15.3 seconds, i.e. more than 300 times faster. Our choice of $\alpha$ corresponds to testing a million independent variants~\citep{peer}, a level that has been widely-adopted as the standard for genome-wide work

We emphasize that this approximation of the Type I error rate is entirely general; any test statistic $T$ can be used, including the familiar Wald, Score, likelihood ratio test statistics (see e.g. \citet{Ma}) or more sophisticated choices such as the Firth test~\citep{Firth,logistf}. Accurate knowledge of these approximate tests enables users to better compare their performance at the nominal $\alpha$.

The formulation of Type I error rate in~(\ref{eqn1}) and its approximation can also directly inform construction of permutation and approximate unconditional tests, as we discuss in Sections~\ref{sec-perm} and~\ref{sec-AU} below. We briefly discuss adjusting for covariates in Section~\ref{sec-adjtest}, for both forms of test.

\subsection{Permutation tests}
\label{sec-perm}

The ability of permutation tests to provide accurate p-values for association testing under minimal assumptions is well-know~\citep{perm1,perm2,perm3,perm4}; where they are applicable, permutation tests are regarded by many analysts as the `gold standard' method. For quantitative traits, a major drawback is that permutations must, in practice, be performed using random number generation~\citep{perm5}. For analysis of binary traits this is not needed; we can instead enumerate all possible permutations and obtain accurate p-values.

%Permutation tests implement choose the nominal ~(\ref{eqn1})  

Using the same notation as above, a permutation test requires an observed test statistic, $T_{obs}$, calculated on the observed data $(m_0,m_1,r_0,r_1)$. We shall consider test statistics from standard score, Wald, likelihood ratio, and Firth test approaches, thus providing a permutation version of them. The test statistic is also calculated for each possible datasets $(m_0,m_1,r_0',r_1')$ obtained by permuting binary outcomes (e.g. case/control labels) among all study subjects, or equivalently permuting the variant/non-variant carrier status among all subjects. Under permutation, the total number of minor allele carriers is the same as in observed data, that is, $r_0+r_1 = r_0'+r_1'$, and under the null hypothesis of no association the probability of observing each dataset follows the hypergeometric distribution~\citep{permbook}:
\[
\tilde{f}(r_0', r_1'; m_0,m_1,r_0+r_1) = {r_0+r_1 \choose r_1'}{m_0+m_1-r_0-r_1 \choose m_1 - r_1'}/{m_0+m_1 \choose m_1}
%\label{eqn2}.
\]
The permutation p-value is then defined as
\[
p_{perm}(r_0, r_1; m_0, m_1) = \sum_{r_0', r_1'} \tilde{f}(r_0', r_1'; m_0,m_1,r_0+r_1)1_{ |T_{r_0',r_1'}| \ge |T_{obs}| },
\]
i.e. the sum of probabilities of datasets with the same number of allele carriers that result in more extreme test statistics than $T_{obs}$. The datasets enumerated in this method are illustrated in Figure 2. 

Permutation tests are exact, in the sense that the observed Type I error rate will always be less than or equal to $\alpha$. This result is well-known and dates back to Fisher~\citep{janssen}. In particular, for the rare variant setting, permutation tests will be fairly conservative. A mathematical explanation is given in the Appendix. 

\begin{figure}[h]
\centering
\includegraphics[width=0.5\textwidth]{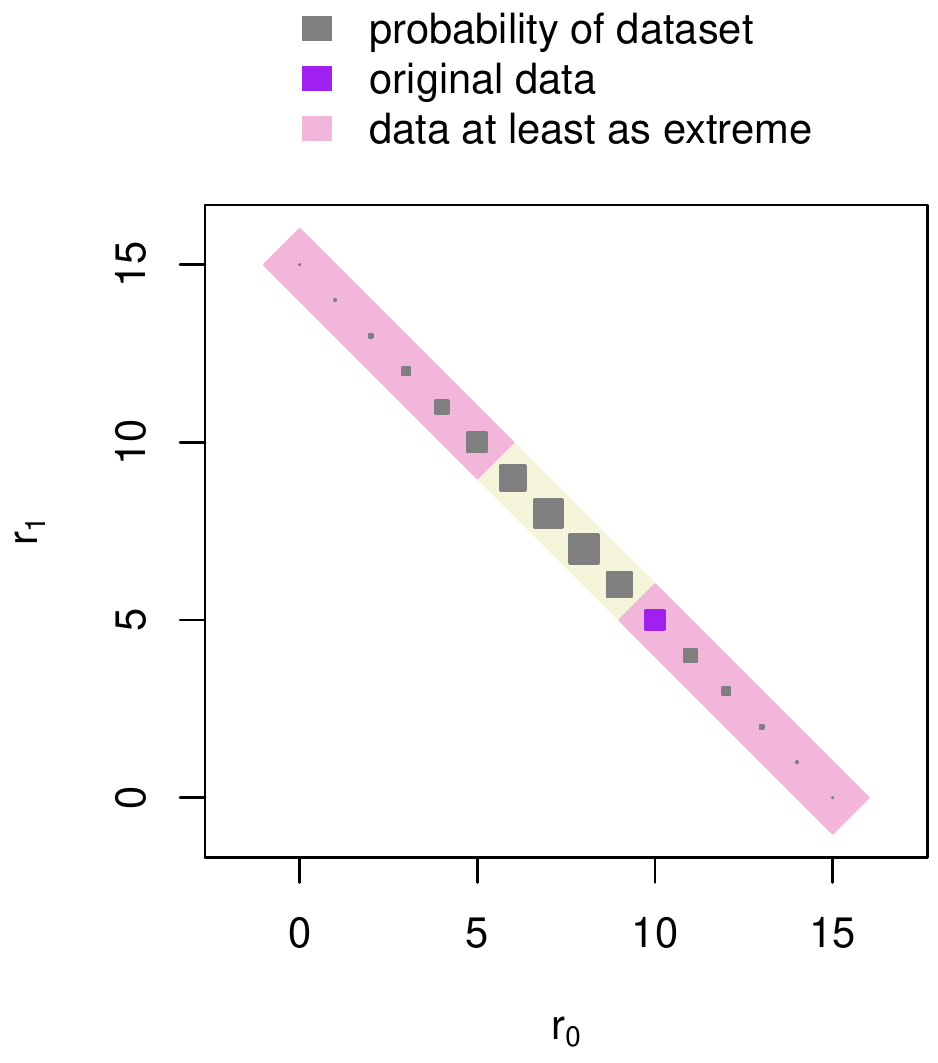}
\caption{Datasets used when calculating the p-value for the permutation version of the Score test, for observed data $(m_0,m_1,r_0,r_1) = (1000,1000,5,10)$. The size of each square corresponds to the probability of observing the corresponding dataset under the null hypothesis. The p-value is represented by the sum of the areas of the squares in the two shaded `tails' of the distribution, containing all datasets with Score test statistic at least as extreme as the observed data.}
\label{fig2}
\end{figure}

\subsection{Approximate Unconditional (AU) tests}
\label{sec-AU}

Approximate Unconditional (AU) tests~\citep{SK:1990} provide Type I error rates closer to the nominal level than permutation approaches. Unlike permutation tests, AU tests are not guaranteed to always strictly control the Type I error rate, but this anti-conservatism (where it occurs at all) is usually very mild.

Using the same notation as above, AU tests calculate a test statistic $T_{obs}$ from the observed data $(m_0,m_1,r_0,r_1)$ and from all possible datasets $(m_0,m_1,r_0',r_1')$ but \textit{without} the restriction that $r_0+r_1 = r_0'+r_1'$. The probability of observing each dataset under the null hypothesis is calculated using fitted binomial distributions, i.e.
\[
\hat{f}(r_0', r_1'; m_0,m_1,r_0,r_1) = {m_0 \choose r_0'}{m_1 \choose r_1'}\bigg(\frac{r_0+r_1}{m_0+m_1}\bigg)^{r_0'+r_1'}\bigg(1 - \frac{r_0+r_1}{m_0+m_1}\bigg)^{m_0 +m_1 - r_0' - r_1'}.
\]
The AU test's p-value is then defined as
\begin{equation}
p_{AU}(r_0, r_1; m_0, m_1) = \sum_{r_0', r_1'} \hat{f}(r_0', r_1'; m_0,m_1,r_0,r_1)1_{ |T_{r_0',r_1'}| \ge |T_{obs}| },
\label{eqn4}
\end{equation}
i.e. the sum of probabilities of datasets that result in more extreme test statistics than $T_{obs}$. We can then apply Equation (\ref{eqn1}) and write the Type I error rate as:
\begin{equation}
T1ER(\alpha) = \sum_{0\leq r_0'\leq m_0, 0\leq r_1'\leq m_1} \hat{f}(r_0', r_1'; m_0, m_1, EMAC)1_{p_{AU}(r_0', r_1'; m_0, m_1) < \alpha},
\end{equation}

Compared to the permutation test, the AU test's p-value sums over many more possible datasets, allowing less crude approximation of the Type I error rate. This comes at the cost of using the same data to fit the null binomial models, and hence losing guaranteed control of the Type I error rate. However, in our setting a bigger practical concern is that taking a na\"{\i}ve approach to calculation in Equation~(\ref{eqn4}) would require $(m_0+1)\times(m_1+1)$ evaluations for each p-value, which may be a burden, as with Equation~(\ref{eqn1}). A much quicker approach that is still adequate in practice uses the same zeroing-out idea as before -- we only sum elements $(r_0', r_1')$ in~(\ref{eqn4}) for values of $r_0'+r_1'$ between the upper and lower $10^{-12}$ quantiles of the $Binom(m_0+m_1, \frac{r_0+r_1}{m_0+m_1})$ distribution. 

\begin{figure}[h]
\centering
a) \raisebox{-3.2in}{\includegraphics[width=0.46\textwidth]{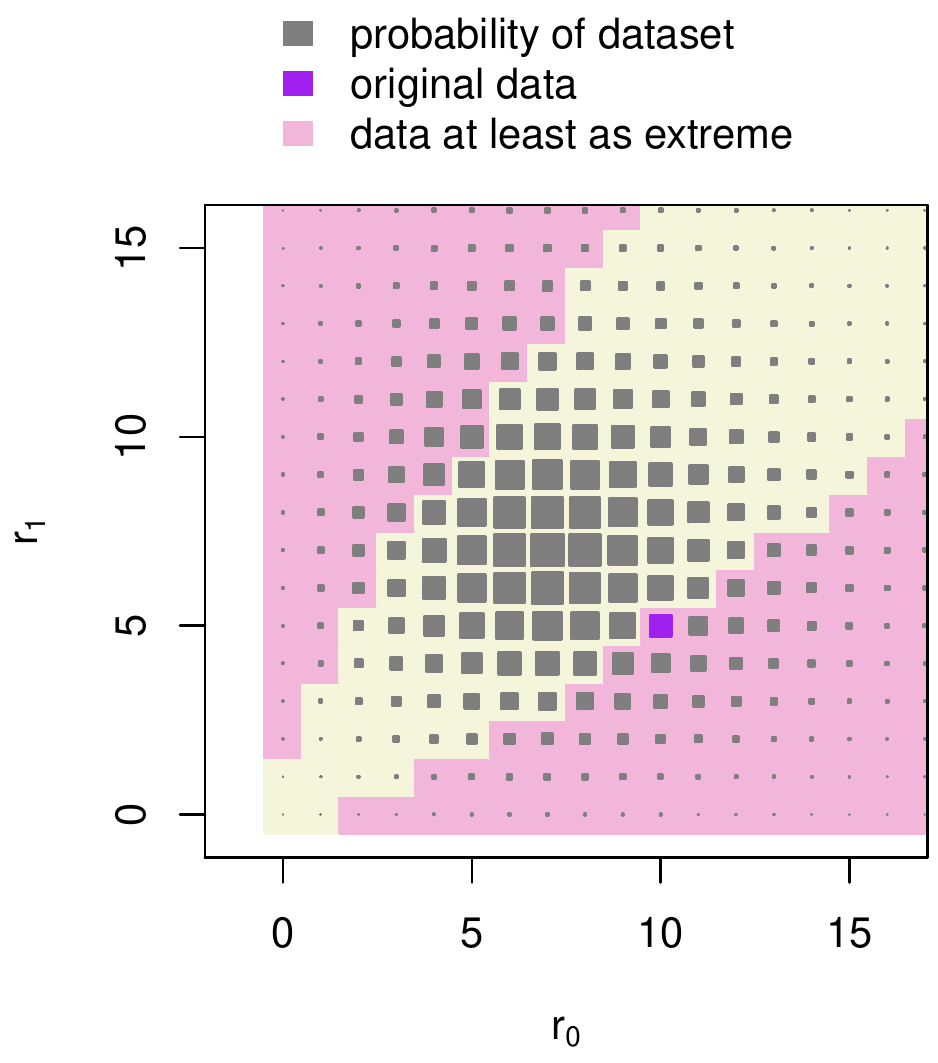}} b) \raisebox{-3.2in}{\includegraphics[width=0.46\textwidth]{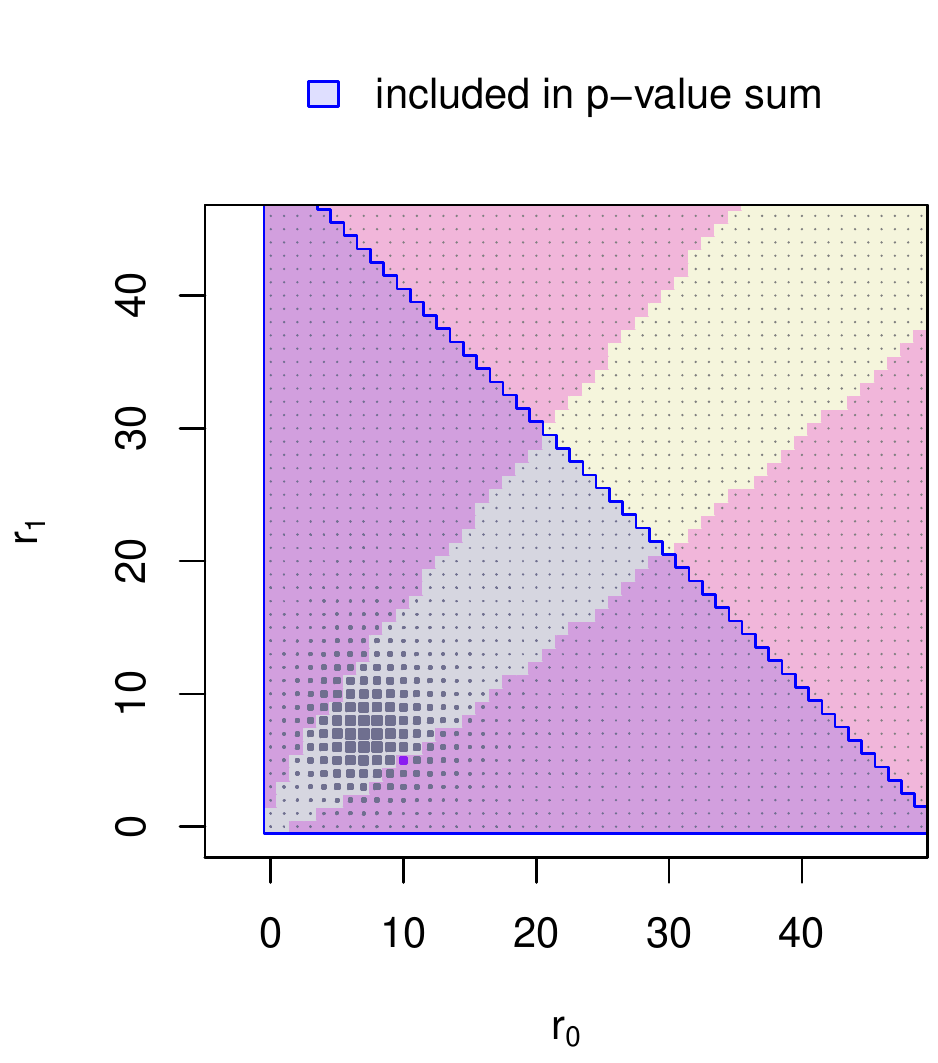}}
\caption{Datasets used when calculating p-value for the AU version of the Score test, for observed data $(m_0,m_1,r_0,r_1) = (1000,1000,5,10)$. The size of each square corresponds to the probability of observing the corresponding dataset under the null hypothesis. In a), the p-value is represented by the sum of the areas of the squares in the two shaded areas, containing all datasets with Score test statistics at least as extreme as the observed data. In b), we show how truncation at the upper $10^{-12}$ quantile of the fitted distribution of $r_0$ and $r_1$ would zero out many datasets, making calculation much quicker.}
\label{fig3}
\end{figure}

The datasets enumerated in this method are illustrated in Figure~\ref{fig3}. As with the calculation of Type I errors in Section~\ref{sec2}, the zeroing out leads to a slight understatement of the p-value compared to complete enumeration. However, understating the p-value by at most $2\times 10^{-12}$ is  a very minor concern when $\alpha=10^{-8}$, several orders of magnitude greater, and comes in return for a substantial speed increase. For example, computing an AU p-value under the Score test with data $(m_0,m_1,r_0,r_1) = (5000,5000,10,50)$ takes 0.05 seconds on a standard laptop with zeroing out and 28.4 seconds without, i.e. over 500 times faster.

The AU approach, like the permutation approach, is completely general, and AU versions of any test can be implemented. We shall use standard Score, Wald, likelihood ratio and Firth tests.

\subsection{Adjusting for covariates}
\label{sec-adjtest}

Both permutation and AU tests permit adjustment for covariates through stratification, i.e. only using information about association from within groups of subjects for whom confounding factors (for example ancestry) are held constant~\citep{clayton}.

Extending the previous notation, for stratified tests we now refer to vectors $\bm_0, \bm_1, \br_0, \br_1$, each of length $q$, where $q$ is the number of strata defined by the levels of one or more categorical covariates. Indexing strata by $i$, with $1\le i \le q$, for each stratum $i$ the stratified test enumerates all possible strata-specific datasets $(m_{0i},m_{1i},r_{0i}',r_{1i}')$ such that $r_{0i}' + r_{1i}' = r_{0i} + r_{1i}$, computing a test statistic for each. The test statistics $T_i(r_{0i}, r_{1i})$ from each strata are combined (by default they are added) to produce a single test statistic for the whole dataset; formally we define
\[
T_{\br_0', \br_1'} = \sum_{i=1}^q  T_i(r_{0i}, r_{1i}).
\]
The p-value, which as before compares this single test statistic to what might have been observed under the null, uses the hypergeometric distribution for each set of stratum-specific counts. We write the probability of observing specific datasets as
\begin{eqnarray*}
\hat{f}(\br_0',\br_1';\bm_0,\bm_1,\br_0,\br_1) &=& \prod_{i=1}^q \hat{f_i}(r_{0i}', r_{1i}'; m_{0i},m_{1i},r_{0i},r_{1i}) \\
\mathrm{where}\,\, \hat{f_i}(r_{0i}', r_{1i}'; m_{0i},m_{1i},r_{0i},r_{1i}) &=& {r_{0i}+r_{1i} \choose r_{1i}'}{m_{0i}+m_{1i}-r_{0i}-r_{1i} \choose m_{1i} - r_{1i}'}/{m_{0i}+m_{1i} \choose m_{1i}},
\end{eqnarray*}
and formally define the p-value as 
\[
p_{strat.perm}(\br_0',\br_1';\bm_0,\bm_1,\br_0,\br_1) = \sum_{r_0', r_1'} \hat{f}(\br_0',\br_1';\bm_0,\bm_1,\br_0,\br_1) 1_{ |T_{\br_0',\br_1'}| \ge |T_{\br_0,\br_1}| },
\]
i.e. the sum of probabilities of datasets that result in more extreme test statistics than $T_{obs}$, where $T_{obs}$ is the test statistic corresponding to the data that was observed.

The stratified AU test is constructed from the same steps as the permutation except for three differences, described earlier in Section~\ref{sec-AU}. First, the datasets considered for each strata include any values of $0\le r_{0i}' \le m_{0i}$ and $0\le r_{1i}' \le m_{1i}$. Second, the probabilities $\hat{f}$ of each dataset are constructed from fitting a null binomial model within each strata. Third, summands within each strata are zeroed-out for which the total contribution is no more than $2 \times 10^{-12}$. 

Our approach removes confounding effects by using stratified analysis. Implemented carefully, there is little to choose between use of stratification versus model-based regression adjustment. In line with Clayton and Hills (1993, Statistical Methods in Epidemiology, pg 273) we find it appealing that the stratification approach forces careful consideration of a which confounders are a priori most important to adjust for, and for stratification approaches to be based closely on the scientific question of interest. Moreover, categorizing confounding into strata is the only approach under which our enumeration approach for exact inference is feasible; regression-based alternatives with continuously-valued covariates and standard computing resources would have to compute p-values by some form of Monte Carlo method, with consequent Monte Carlo error and long compute times. 

\section{Analytical calculation results}
\label{sec3}

To illustrate analytical calculations, we set the total sample size to be $N = 10,000$---close to that seen in Section~\ref{sec4}'s example---and considered case:control matching ratios of 1:1, 1:3, and 1:19. We set the nominal significance level at $\alpha = 5 \times 10^{-8}$, and use $EMAC$ ranging from 1 to 100.

Permutation and AU versions of Score, Wald, likelihood ratio and Firth tests were examined. For comparison we also computed the standard Score, Wald, likelihood ratio, Firth tests, and Fisher's exact test, which is itself a permutation test. For permutation, AU, and standard tests we also considered a regularized Wald test, which avoids undefined test statistics by adding 0.5 to each cell count when any count is zero. 

\begin{figure}[h]
\centering
\includegraphics[width=0.98\textwidth]{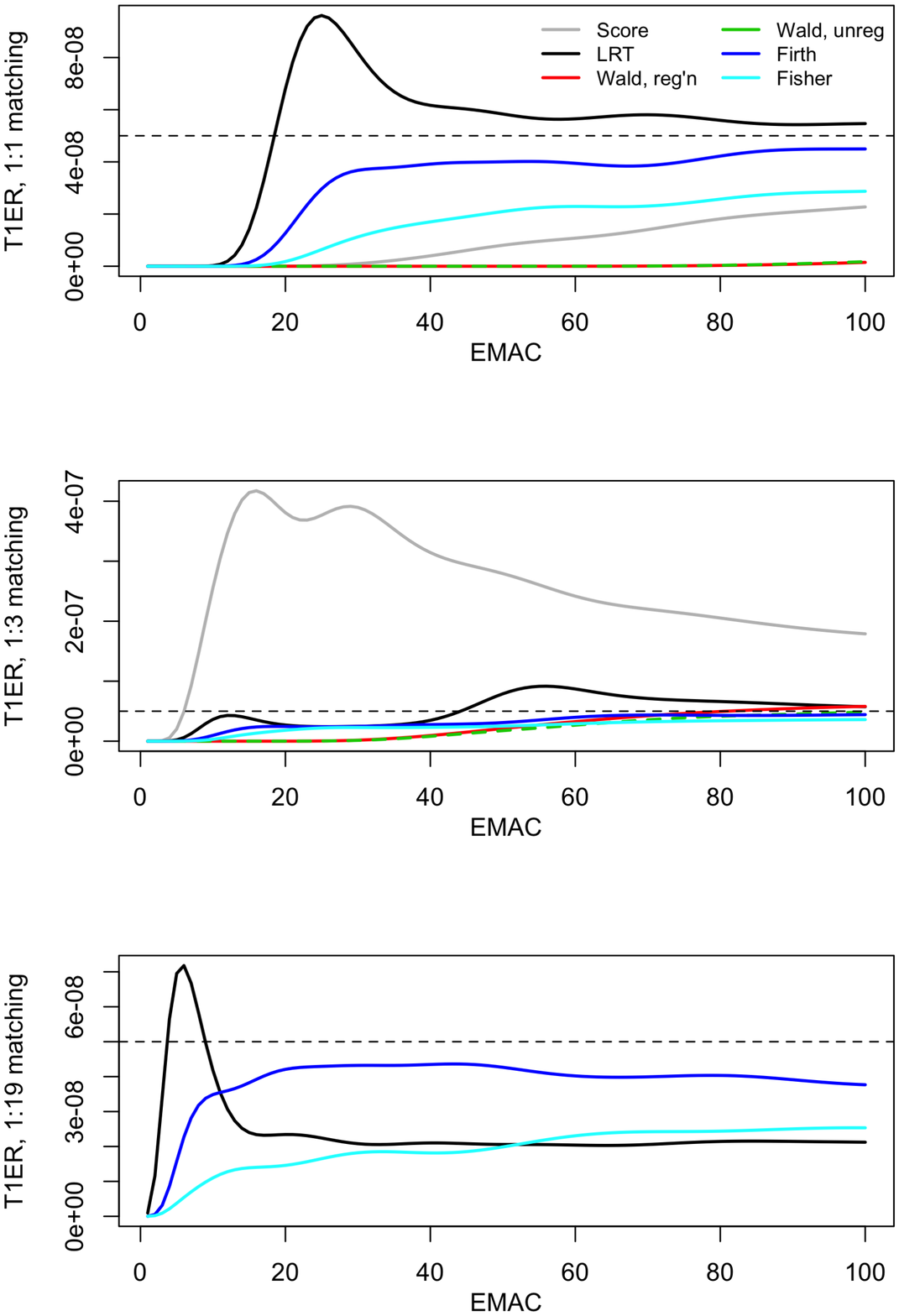}
\caption{$T1ER$ versus EMAC for various standard tests at $\alpha=5\times10^{-8}$ (dotted line) and total N=10,000, grouped by matching ratio. All calculations use the zeroing-out technique of Section~\ref{sec2}, and so understate the true $T1ER$ by no more than $10^{-12}$. Score and Wald tests are omitted from the final plot due to gross violation of the nominal $\alpha$.}
\label{fig4}
\end{figure}

As seen in Figure~\ref{fig4}, the tests based on standard asymptotics do not adequately control the Type I error rate. In the balanced design, the tests are overly conservative, with the exception of the likelihood ratio test, which is anti-conservative. The Score test has very a large Type I error rate under the 1:3 ratio, so is presented separately. This is also true under the 1:19 ratio for the Score and Wald tests, which are omitted. The other tests continue to be conservative, and the likelihood ratio test's Type I error rate is too large over certain ranges. The Firth test consistently performs the best. 

\begin{figure}[h]
\centering
\includegraphics[width=0.98\textwidth]{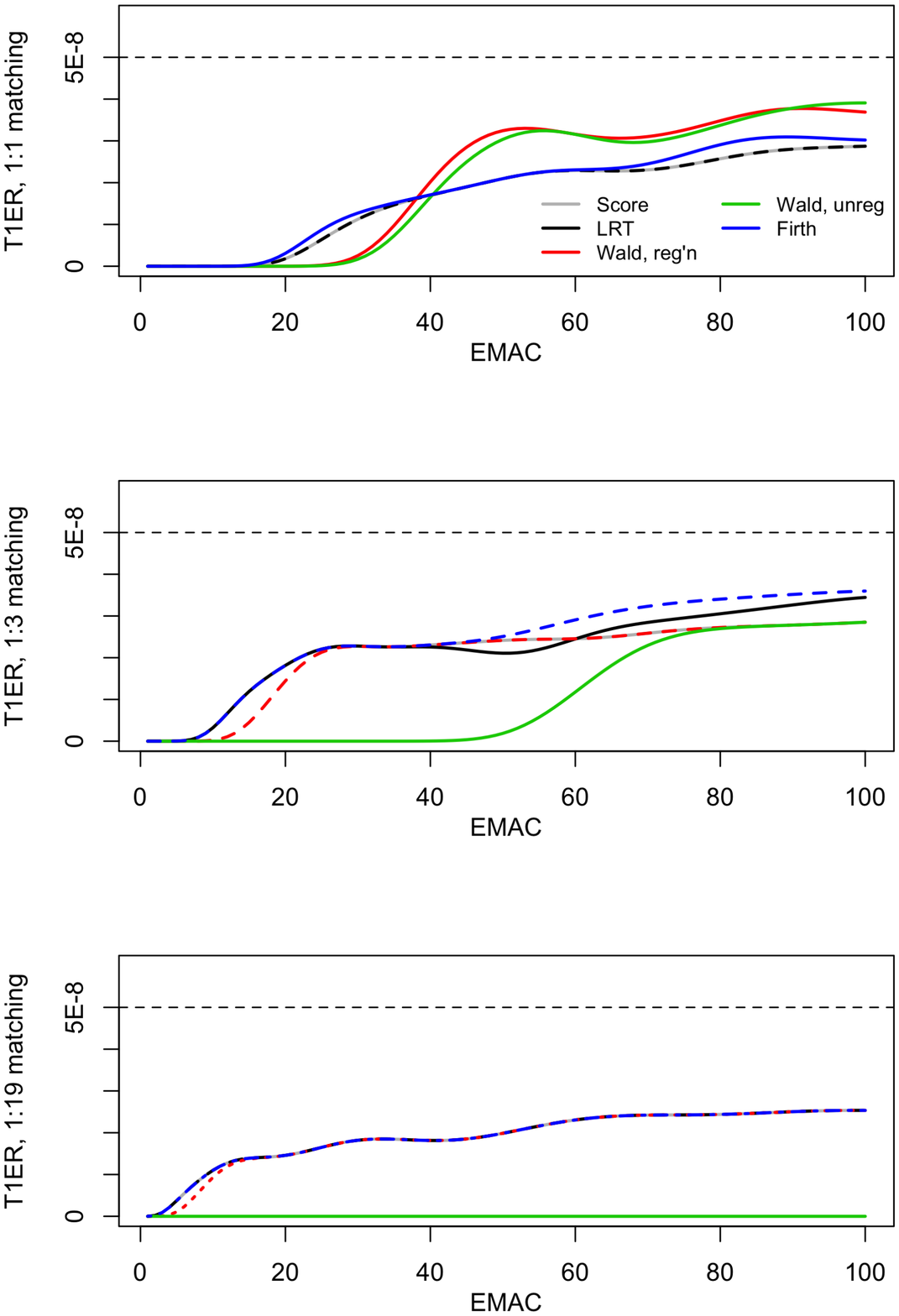}
\caption{$T1ER$ versus EMAC for permutation versions of standard tests as described in Section~\ref{sec-perm} at $\alpha=5\times10^{-8}$ (dotted line) and total N=10,000, grouped by matching ratio. All $T1ER$ rate calculations use the zeroing-out technique of Section~\ref{sec2}, and so understate the true $T1ER$ by no more than $10^{-12}$.}
\label{fig5}
\end{figure}

In Figure 5, we see that the permutation tests improve upon most of the standard tests, though remain more conservative than the regular Firth test. While these tests have the advantage of being exact, as the case:control ratio becomes more unbalanced, the Type I error rate becomes more conservative. Under the 1:19 ratio, all tests perform nearly identically, with the exception of the unregularized Wald test.

\begin{figure}[h]
\centering
\includegraphics[width=0.98\textwidth]{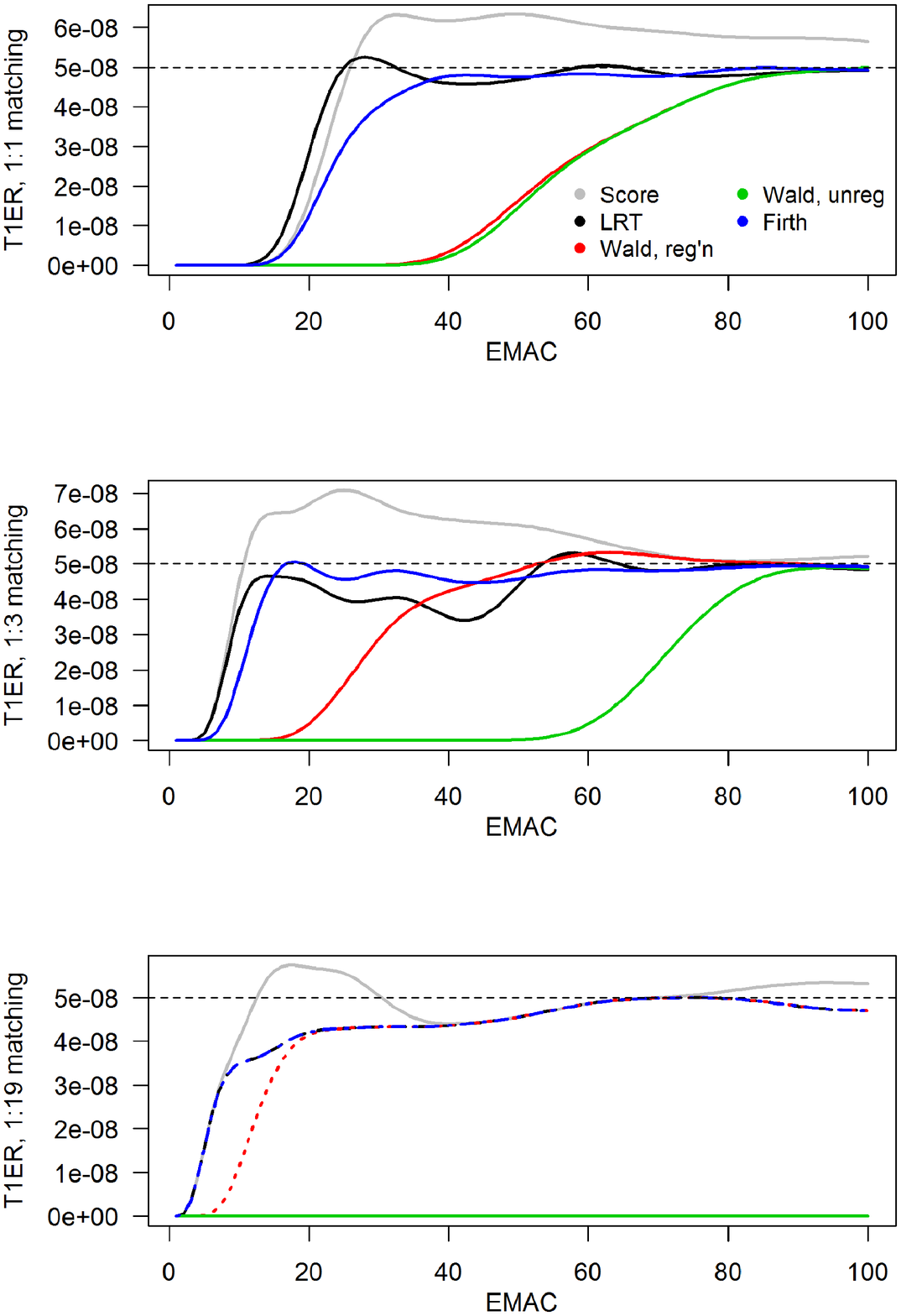}
\caption{$T1ER$ versus EMAC, for AU versions of standard tests as described in Section~\ref{sec-AU}, at $\alpha=5\times10^{-8}$ (dotted line) and total N=10,000, grouped by matching ratio. All $T1ER$ rate calculations use the zeroing-out technique of Section~\ref{sec2}, and so understate the true $T1ER$ by no more than $10^{-12}$.}
\label{fig6}
\end{figure}

In Figure 6, we see that the AU tests show a large improvement over standard and permutation tests, especially in the AU likelihood ratio and AU Firth tests. Though they are not exact, the excess Type I error rate is mild. Note that under the 1:19 ratio, the Firth and likelihood ratio tests perform identically. 

\section{Application: Rhabdomyolysis case-control study}
\label{sec4}

The data comes from an exome-sequencing study, in which 9,763 subjects who used statins were considered; 211 cases with rhabdomyolysis and 9,552 controls. The rationale for this design are described in detail by~\citet{marciante2011cerivastatin}. Our interest was primarily in assessing if there existed rare genetic variants associated with developing rhabdomyolysis in statin users. We defined `rare' variants as those where less than or equal to 100 study participants carried the minor allele. Variants with less than 5 minor allele carriers were also removed, as these provide no ability to produce significant values at the low $\alpha$ threshold used in this form of study. Finally, for quality control, we filtered out variants with a genotyping rate of less than 0.85. Applying these filters left 161,428 variants, and there are no covariates for which to adjust in this analysis. 

\begin{figure}[h!]
\centering
\includegraphics[width=0.85\textwidth]{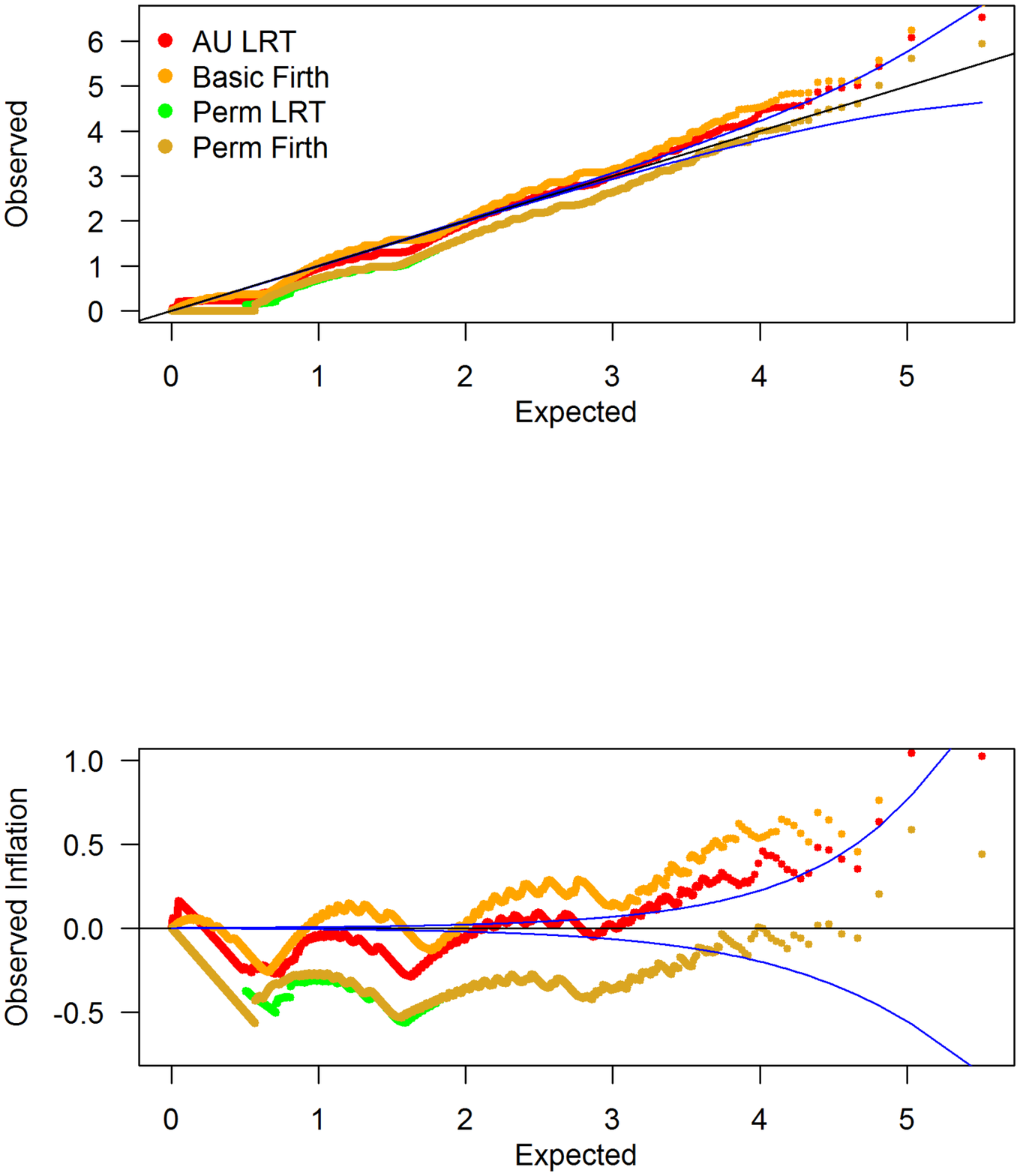}
\caption{QQ plot and 45 degree rotated QQ plot of -log10 p-values for rhabdomyolysis dataset, as described in Section~\ref{sec4}. After quality control filtering, 161,428 variants are analyzed, with between 5 and 100 minor allele carriers each. For each method, the QQ plot shows the ordered p-values versus the corresponding expected value from null, i.e. Uniform(0,1) p-values. The blue cone shape indicates pointwise 95\% prediction bounds for each ordered p-value. The rotated plot shows the same results, but where the y-axis shows the -log10 observed p-value minus the -log10 expected p-value; the blue cone has the same interpretation as before.}
\label{fig7}
\end{figure}

We applied the AU and permutation versions of the likelihood ratio test, and the permutation and standard Firth test to all variants. The entire analysis took approximately 6.5 hours on a shared server, using a single CPU. The AU version of the Firth test was not used due to its high computational burden. The resulting QQ plot and a plot of the inflation (45 degree rotated QQ plot) observed are given in Figure~\ref{fig7}.  

While some granularity in the larger p-values is present on the left hand of both plots, based on our numerical results, it is reasonable to expect that the AU likelihood ratio test provides the best control of the Type I error rate. Applied to this dataset, we observe that the AU likelihood ratio test results in significantly less inflation than the standard Firth test. Therefore, we believe that the right-hand tail of variants with p-values declared to be significant under the AU test are more accurate statements of statistical significance than the other methods.

\section{Discussion}
\label{sec5}

We have developed and implemented association tests for rare genetic variants, that control Type I error rates better than standard asymptotic tests. Of the tests proposed in this paper, the AU version of the likelihood ratio and Firth tests perform the best, particularly when the ratio of cases to controls is extreme. However, the AU version of the Firth test has a notably higher computational burden than competitors. Therefore, we recommend the the AU likelihood ratio test, for large genome-wide studies. If an exact test is necessary, then a permutation test is recommended; though conservative, it tends to show an improvement over standard tests. We note that if the expected number of minor allele carriers is less than 20, then no test will perform adequately, and conservative control of the Type I error rate is the best achievable property. 

The methods described here have been implemented in an R package, \texttt{AUtests}. This package contains the functions \texttt{basic.tests}, \texttt{perm.tests}, and \texttt{au.tests}, which implement all the respective standard, permutation, and AU tests for a given vector of counts $(m_0,m_1,r_0,r_1)$, returning a vector of p-values. The AU Firth test is implemented in a separate function, \texttt{au.firth}, due to its increased computational time. For a typical dataset $(m_0,m_1,r_0,r_1)=(10000,10000,50,50)$, on a standard laptop, the \texttt{basic.tests} function takes 0.03 seconds of CPU time, the \texttt{perm.tests} function takes 0.21 seconds, the \texttt{au.tests} function takes 0.39 seconds, and the \texttt{au.firth} function takes 51 seconds. To account for covariates, appropriately categorized, the package also contains the functions \texttt{au.test.strat} and \texttt{perm.test.strat}, which implement stratified AU and permutation likelihood ratio tests. The package is available on CRAN.

\section{Acknowledgments}

Research reported in this paper was supported by the National Institute on Aging of the National Institutes of Health under award numbers U01AG049505 and U01AG049507, and by the National Heart, Lung, and Blood Institute of the National Institutes of Health under award number R01 HL078888. The content is solely the responsibility of the authors and does not necessarily represent the official views of the National Institutes of Health.

\clearpage

\appendix 

\section{Appendix: exact control of permutation tests}

In this appendix, we show that permutation tests give exact control of the Type I error rate. Rewriting equation~(\ref{eqn1}) as a double summation over the observed minor allele count ($t := r_0+r_1$) and the number of these in the controls, we obtain
\[
T1ER(\alpha) = \sum_{0\le t \le m_0+m_1}g(t ;m_0,m_1) \sum_{r_0}f(r_0; m_0, m_1, t)1_{p(r_0, r_1; m_0, m_1) < \alpha}
\]
where $g()$ denotes the probability of the observed minor allele count, and $f()$ gives the probability of the observed counts in cases and controls given the minor allele count $r_0+r_1$ --- so $f()$ supports values of $r_0$ between  $\max(0, t - m_1)$ and $\min(m_0,t)$. 

By construction, the inner sum always gives a value less than or equal to $\alpha$; the outer sum averages these, and so is similarly bounded. However, particularly for rare variants, the inner sum considers a small set of possible permutations, as illustrated in Figure~\ref{fig2}. While this makes the test fast enough that zeroing-out is not required, it means that for small $\alpha$, the actual Type I error rate, while below $\alpha$, will be quite conservative for many values of $m_0$ and $m_1$.

\section{Appendix: AU test power calculations}

In this section, we show power calculations for the AU Firth test, under the same scenarios considered in the main paper. We observe that power decreases as case:control matching ratios become more skewed. In particular, the extremely unbalanced 1:19 ratio requires a very large association in order to have meaningful power, even at higher minor allele counts. 

\begin{figure}[h!]
\centering
\includegraphics[width=0.98\textwidth]{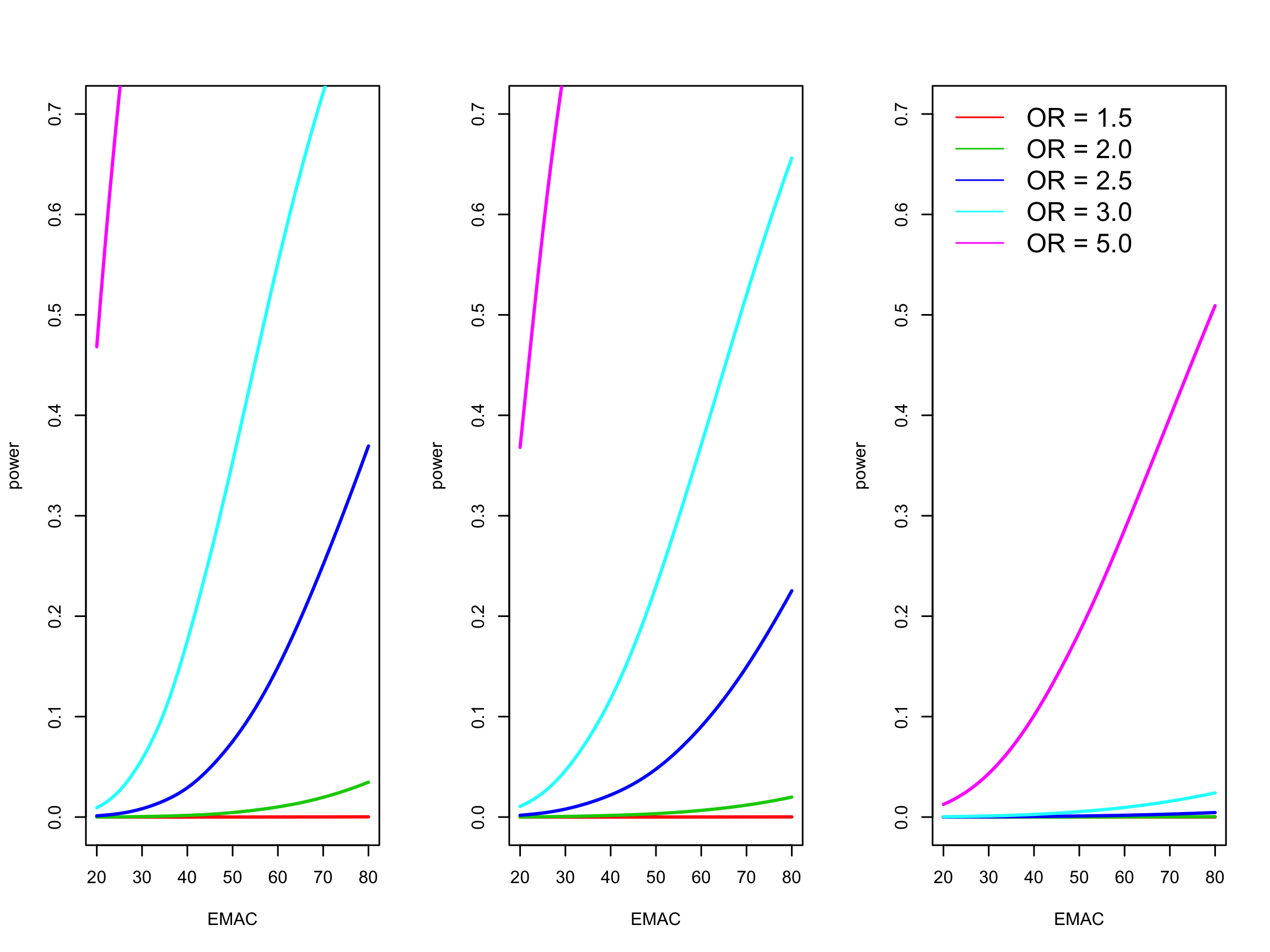}
\caption{Power curves giving the probability of rejecting the null hypothesis of independence by expected minor allele count. Different curves correspond to different odds ratios. Each panel corresponds to a different case:control matching ratio with an overall sample size of 20,000. Left: 1:1, middle: 1:3, right: 1:19 ratio}
\label{power}
\end{figure}

\clearpage

\bibliographystyle{apalike}
\bibliography{manuscript_v9}

\end{document}